\documentclass[12pt]{article}

\usepackage{graphicx}
\usepackage{epstopdf}
\usepackage[centertags]{amsmath}
\usepackage{amssymb}
\usepackage{amsthm}
\usepackage{amsfonts}
\usepackage{ccaption}
\usepackage[usenames]{color}
\usepackage[
      colorlinks=true,
      linkcolor=blue,  
      urlcolor=blue,    
      filecolor=blue,  
      citecolor=black,   
      pdfstartview=FitV,
      pdftitle={},
        pdfauthor={Chris Herzog, Mukund Rangamani,Simon Ross},
        pdfsubject={},
        pdfkeywords={},
        pdfpagemode=None,
        bookmarksopen=true    
      ]{hyperref}
       
\usepackage{rotating}

\makeatletter
\@addtoreset{equation}{section}

\makeatletter
\renewcommand\section{\@startsection {section}{1}{\z@}%
                                   {-3.5ex \@plus -1ex \@minus -.2ex}
                                   {2.3ex \@plus.2ex}%
                                   {\normalfont\large\bfseries}}
\renewcommand\subsection{\@startsection{subsection}{2}{\z@}%
                                     {-3.25ex\@plus -1ex \@minus -.2ex}%
                                     {1.5ex \@plus .2ex}%
                                     {\normalfont\bfseries}}

  \captionnamefont{\bfseries}
  \captiontitlefont{\small\sffamily}
  \captiondelim{: }
  \hangcaption

\def\baselinestretch{1.2}
\newcommand{\be}{\begin{equation}}
\newcommand{\ee}{\end{equation}}
\newcommand{\beq}{\begin{eqnarray}}
\newcommand{\eeq}{\end{eqnarray}}

\def\sec#1{\S \ref{#1}}

\def\req#1{(\ref{#1})}
\def\App#1{Appendix \ref{#1}}

\def\thus{\Longrightarrow}

\def\CN{{\cal N}}
\def\CO{{\cal O}}

\def\CQ{{\cal Q}}

\def\R{{\bf R}}
\def\Sp{{\bf S}}

\def\A5S5{{\rm AdS}_5 \times \S^5}

\def\ord#1{\[ \, #1 \, \]}

\def\vev#1{\langle\, #1 \, \rangle}

\def\ord#1{\CO\left(#1\right)}

\def\AdS#1{AdS$_{#1}$}
\def\q{\gamma^2}
\def\SS{{\mathcal S}}

\title{{\bf \Large Heating up  Galilean holography}}
\date{July 7, 2008}

\author{
Christopher P. Herzog$^a$\footnote{cpherzog@princeton.edu}, \ Mukund Rangamani$^b$\footnote{mukund.rangamani@durham.ac.uk}, \ and Simon F. Ross$^b$\footnote{s.f.ross@durham.ac.uk} \\ \\
\small\sl $^a$ Department of Physics, Princeton University, Princeton, NJ 08544, USA\\
\small \sl $^b$  Centre for Particle Theory \& Department of
Mathematical Sciences,
\\[-1.5mm]
\small \sl Science Laboratories, South Road, Durham DH1 3LE, United Kingdom. \\
}

\begin{document}
\setlength{\baselineskip}{16pt}
\begin{titlepage}
\maketitle
\begin{picture}(0,0)(0,0)
\put(350, 300){PUPT-2274} 
\put(350,285){DCPT-08/39}
\end{picture}
\vspace{-36pt}

\begin{abstract}
 We embed a holographic description of a quantum field theory with
  Galilean conformal invariance in string theory.  The key observation
  is that such field theories may be realized as conventional
  superconformal field theories with a known string theory embedding,
  twisted by the R-symmetry in a light-like direction.  Using the Null
  Melvin Twist, we construct the appropriate dual geometry and its
  non-extremal generalization.  From the nonzero temperature solution
  we determine the equation of state. We also discuss the hydrodynamic
  regime of these non-relativistic plasmas and show that the shear
  viscosity to entropy density ratio takes the universal value $\eta/s
  = 1/4\pi$ typical of strongly interacting field theories with
  gravity duals.
 \end{abstract}
\thispagestyle{empty}
\setcounter{page}{0}
\end{titlepage}

\renewcommand{\baselinestretch}{1.4}  
\renewcommand{\thefootnote}{\arabic{footnote}}


\section{Introduction}
\label{intro}

The AdS/CFT correspondence \cite{Maldacena:1997re, Gubser:1998bc,
  Witten:1998qj} maps relativistic conformal field theories
holographically to gravitational (or stringy) dynamics in a higher
dimensional asymptotically Anti-de Sitter spacetime.  As such, the
correspondence is an important tool for modeling the behavior of
strongly interacting field theories, as the dynamics on the field
theory side is mapped to classical string and gravitational dynamics
in the dual description. Indeed, among other achievements, this
strong--weak coupling duality has improved, at least at a qualitative
level, our understanding of real-time dynamics and transport
properties of the quark-gluon plasma in QCD.

More recently, these holographic ideas have been applied to conformal field
theories arising from condensed matter systems. There is a large class
of interesting strongly correlated electron and atomic
systems that can be created and studied in
table-top experiments. In special cases, these systems exhibit relativistic
dispersion relations, and the dynamics near a critical point
is then well described by a relativistic conformal field theory. 
It is precisely such field theories which may be studied using holographically dual AdS
geometries.  Recent work 
(see refs.\ \cite{Herzog:2007ij}
for a sampling)
has already applied this AdS/CMat correspondence to strongly
correlated electrons, superconductors, the quantum Hall effect, and more. 

More ambitiously, one can ask whether the holographic approach can be
extended to the non-relativistic theories describing most condensed
matter systems. In particular, can field
theories with Galilean scaling symmetry (see
refs.\ \cite{Hagen:1972pd,Mehen:1999nd,Nishida:2007pj} for discussions of
non-relativistic CFTs) have a holographic dual? Just as the
Poincar\'e algebra can be extended to the conformal algebra in
relativistic quantum field theories, one can extend
the Galilean algebra (symmetry of non-relativistic field theories) to
the so called Schr\"odinger algebra \cite{Hagen:1972pd}.  
Fermions at unitarity are conjectured to realize this Schr\"odinger symmetry.
The scale invariance is achieved by fine tuning the fermions --- with an external
magnetic field for example ---  to obtain a massless bound state, thus making the scattering length effectively infinite.
 These systems are of increasing interest in the
context of trapped cold atoms at a Feshbach resonance. Indeed refs.\  
\cite{Gelman:2004fj,Schafer:2007ib,Schafer:2007xw,Rupak:2007vp} claim 
that these cold atom systems provide another example of a
nearly ideal fluid with very low viscosity, like the quark-gluon plasma. See refs.\ \cite{OHara:2002vh,Bartenstein:2004rf,Kinast:2004ts,Bourdel:2004bl,Turlapov:2008ss} for experimental studies of cold atom systems and their hydrodynamic transport coefficients. Indeed the latest result of 
 ref.\ \cite{Turlapov:2008ss} predicts a value of $\eta/s$ slightly above the values obtained for the quark-gluon plasma in heavy-ion collisions, putting it well into the strongly coupled regime. Having a holographic description
would certainly be helpful for understanding the strongly coupled
dynamics one encounters in these systems.

Important steps in this direction were taken in
refs.\ \cite{Son:2008ye,Balasubramanian:2008dm}, where gravitational
backgrounds dual to non-relativistic conformal field theories were
proposed.\footnote{%
See also refs.\ \cite{horvathy} for related work.
}
 These dual geometries involve a pp-wave deformation of
AdS. In this paper, we will continue the exploration of the bulk
geometry proposed as the dual of non-relativistic conformal field
theories. We will show how the geometry can be realized in a string
theory context, and discuss non-extremal generalizations, dual to
non-relativistic conformal field theories at nonzero temperature. A
slightly different duality involving pure AdS bulk spacetimes was
proposed in refs.\ \cite{Goldberger:2008vg,Barbon:2008bg}. In the next
section, we will review the proposed duality of
refs.\ \cite{Son:2008ye,Balasubramanian:2008dm}, and consider the advantages
and disadvantages compared to the rival proposals of
refs.\ \cite{Goldberger:2008vg,Barbon:2008bg}.\footnote{%
See also ref.\ \cite{Wen:2008hi} for related work.
}

In \sec{solgen} we will show how the geometries of
refs.\ \cite{Son:2008ye,Balasubramanian:2008dm} can arise in string theory,
and construct nonzero-temperature generalizations of them. We describe
how the solutions can be constructed by the Null Melvin Twist
\cite{Alishahiha:2003ru,Gimon:2003xk}, which was originally invented
to construct asymptotically plane wave black holes. For the case of
$d=2$ spatial dimensions the conformally invariant theory will be
realized as the world-volume theory on D3-branes with a light-like
twist in the R-symmetry directions. Starting from $\CN =4$ Super-Yang
Mills (SYM) on the D3-brane world volume, the effect of the twist can
be understood as adding a dimension five Lorentz violating operator,
which deforms the asymptotic AdS geometry to the desired pp-wave form.
The twist will break the R-symmetry of $\CN =4$ SYM from $SU(4)$
down to $SU(3) \times U(1)$.\footnote{There are twists of the R-symmetry that preserve as many as 8 supercharges \cite{Alishahiha:2003ru}. We use a simple twist which does not preserve any supersymmetry because of the resulting form of the $H_{(3)}$ flux.} We thus realize a non-relativistic conformal field theory directly in terms of a discrete light-cone
quantization (DLCQ) of a deformation of $\CN =4$ SYM. Such theories were discussed previously in \cite{Bergman:2000cw,Alishahiha:2003ru} and belong to a class of non-local field theories called dipole theories.\footnote{%
We would like to thank Allan Adams for mentioning this DLCQ interpretation in informal discussion.
}

Another interesting class of non-relativistic field theories which
arise in string theory are a special case of non-commutative field
theories; in these theories the geometry is supported by fluxes which
break the spatial rotational symmetries.  These do not have the
Schr\"odinger symmetry but obey a generalized scaling symmetry. The
simplest such model is the light-like non-commutative $\CN=4$ SYM
described originally in ref.\ \cite{Aharony:2000gz}, whose holographic dual
was constructed in ref.\ \cite{Alishahiha:2000pu}. These geometries were
investigated for their causality properties in
refs.\ \cite{Hubeny:2005qu,Hubeny:2005pz}, where the Galilean structures were
naturally shown to arise; we will discuss this issue of causal
structure further in \sec{rev} and some specific examples in \App{fluxbgs}.

As the spacetime geometries we consider are constructed by the Null
Melvin Twist solution generating technique, it is easy to construct
the non-extremal versions of the solutions considered in refs.\ 
\cite{Son:2008ye,Balasubramanian:2008dm}. In \sec{therdof}, we give a
preliminary consideration of the thermodynamics of the non-extremal
geometries.  We argue that the black hole solution which is dual to
the thermal version of the twisted D3-brane theory corresponds to
working in a grand canonical ensemble with chemical potential for the
particle number (realized as momentum in the light-cone
direction). Given this interpretation, we discuss how to obtain this
grand canonical partition function via a Euclidean quantum gravity
saddle point computation.
We then undertake a detailed investigation of the asymptotics and
action for the spacetimes of interest in section
\sec{asymcons}, recovering from this analysis the complete
thermodynamics of the dual non-relativistic field theory. We then turn
to the hydrodynamic description of the non-relativistic conformal
plasmas and calculate the shear-viscosity in \sec{shear}. We find that
$\eta/s$ has the universal value
$1/4\pi$ typical of strongly interacting field theories with gravity duals 
\cite{Kovtun:2004de}. We conclude with a discussion in \sec{discuss}. In
\App{fluxbgs} we discuss how to obtain metrics of the form
\req{vacmet} supported by $p$-form fluxes that break rotational invariance
and their realization in
string theory.

\noindent
{\bf Note Added:}
 In concurrent work Adams, Balasubramanian, and McGreevy
\cite{MITgroup}  and Maldacena, Martelli and Tachikawa \cite{Juan} have obtained very similar results to ours. We would
like to thank Allan Adams for extensive discussions regarding their
results during the BIRS workshop.

\section{The geometry dual to Galilean CFTs}
\label{rev}

We begin with a brief review of the proposed holographic duality for
non-relativistic field theories of
refs.\ \cite{Son:2008ye,Balasubramanian:2008dm,Goldberger:2008vg,Barbon:2008bg}.
In these proposals, the non-relativistic conformal symmetry is
realized as a subset of a relativistic conformal symmetry with an
additional dimension. The Schr\"odinger algebra is obtained from the
relativistic conformal algebra by reducing along a light-cone. The procedure is
similar to light-cone quantization, where at
fixed light-cone momentum only a Galilean subgroup of the Lorentz
group is manifest.

The holographic dual of a $d$ spatial dimensional Galilean CFT is then
a gravitational solution in $d+3$ dimensions. This dual spacetime
should realize the Galilean scaling symmetry as an isometry. In
ref.\ \cite{Son:2008ye,Balasubramanian:2008dm}, this spacetime is taken to have
a metric of the form\footnote{We consider here $\nu \neq 0$, since $\nu =0$ is
just \AdS{d+2}.}
\begin{equation}
ds^2 = r^2\, \left(-2 \, du\, dv - r^{2 \nu}\, du^2 + d{\bf x}^2 \right) + \frac{dr^2}{r^2},
\label{vacmet}
\end{equation}	
where ${\bf x} = \{x_1 \, \cdots x_d\}$ are the spatial coordinates of
the Galilean field theory.  The light-cone coordinate $u$ is the
boundary time coordinate: the field theory Hamiltonian is conjugate to
$\frac{\partial}{\partial u}$. The role of the
$v$-direction is unclear; it is proposed that we treat this as a
compact direction, in the spirit of DLCQ.  As in AdS/CFT, the bulk
coordinate $r$ should correspond to scale size in the boundary field
theory. The Galilean scaling symmetry is realized as 
\begin{equation}
{\bf x } \sim \lambda \, {\bf x}, \quad u \sim \lambda^{\nu +1}\, u , \quad v \sim \lambda^{1-\nu} \, v , \quad r \sim \lambda^{-1} \, r.
\label{galscaling}
\end{equation}	
In the special case of $\nu =1$ it is expected that the scaling
symmetry extends to full Galilean conformal invariance, realizing the
Schr\"odinger algebra. In this special case, $v$ is invariant under
scaling; the Galilean scale invariance requires that the time
coordinate, $u$, scales twice as fast as the spatial coordinates ${\bf
  x}$. For details of the Schr\"odinger algebra we refer the reader to
ref.\ \cite{Nishida:2007pj}. We will primarily focus on the $\nu =1$ case,
but will also mention realizations of the $\nu =2$ case in terms of
non-commutative field theories in \App{fluxbgs}.

The geometry \req{vacmet} is a solution to Einstein's equations with
negative cosmological constant, with matter whose stress tensor is of
the null dust form $T_{uu} \propto r^{2\nu+2}$.  Ref.\ \cite{Son:2008ye}
modeled the matter using a massive vector field, while
ref.\ \cite{Balasubramanian:2008dm} used an Abelian-Higgs action.  
We will discuss below how to embed the construction of
ref.\ \cite{Son:2008ye} into Type IIB string theory and realize the line element
\req{vacmet} as the near-horizon geometry of a twisted D3-brane solution.

On the other hand, refs.\ \cite{Goldberger:2008vg,Barbon:2008bg}
proposed that the dual geometry can be pure AdS, with the
relativistic conformal symmetry broken to Galilean symmetry simply by
compactification of the $v$ coordinate, which singles out a
preferred light-cone direction. Such a modification of AdS would certainly be 
a simpler
setting in which to study Galilean symmetry, but we feel that the
geometry \eqref{vacmet} is a more natural dual for such
non-relativistic conformal theories.

The main reason we prefer the original proposal of 
refs. \cite{Son:2008ye,Balasubramanian:2008dm}
 to the simplified proposal
of refs.\ \cite{Goldberger:2008vg,Barbon:2008bg} 
is that the causal structure of \eqref{vacmet}
naturally reproduces the Galilean light cone of the field theory. The
causal structure of a non-relativistic field theory is degenerate ---
interactions can propagate instantaneously. 
While a bulk geometry with a well-behaved causal structure
cannot be holographically dual to a non-relativistic field
theory, the spacetime geometry \req{vacmet} evades this issue
beautifully --- its causal structure is also degenerate and in such a way
as to be consistent with the boundary Galilean
invariance.\footnote{The special case $\nu =0$ is of course pure AdS
  with a well-behaved causal structure.}

The spacetime \req{vacmet} is conformal (with an overall conformal
factor $r^2$) to a pp-wave spacetime, and this
pp-wave spacetime is known to be non-distinguishing \cite{Flores:2002fx,Hubeny:2003sj}. Non-distinguishing means that while the spacetime \req{vacmet} is
causal (in the sense of not having closed causal curves), there are
distinct points in the spacetime which have identical past and future
sets,\footnote{The timelike future $I^+(p)$ for a point $p$ is the set
  of points which can be reached from $p$ by future-directed timelike
  curves; timelike past is defined similarly. Causal future/past are defined likewise in terms of causal (timelike or null) curves.} thereby preventing us from distinguishing spacetime
events by reference to their past and future sets.  In fact, in
\eqref{vacmet}, all points on a surface with $u = u_0$ (and arbitrary
values of other coordinates) have an identical causal future/past
\cite{Hubeny:2003sj}. But Galilean CFTs have precisely such a causal
structure; all spatial points on an equal time surface can influence
any arbitrary spatial point at an infinitesimal time later.\footnote{%
  Note that refs.\ \cite{Hubeny:2005qu,Hubeny:2005pz}, in considering
  the holographic dual of non-commutative $\CN =4$ Super-Yang Mills
  with light-like non-commutativity \cite{Alishahiha:2000pu}, have
  already studied the causal properties of precisely this $\nu = 2$
  geometry.  }  By contrast, a pure AdS spacetime with boundary
conditions engineered to give Galilean invariance does not possess a
bulk light-cone which agrees with the light-cone of the relativistic
field theory.

This consistency of the bulk spacetime causal structure with the
boundary causal structure is a crucial ingredient in the AdS/CFT
correspondence.  Without this agreement, we would easily be able to set
up gedanken experiments where bulk physics and boundary physics
would not agree. Consider for example the question of the singularity structure of the boundary correlation function as discussed in ref.\ \cite{Hubeny:2006yu}; in pure AdS the correlation functions will have a singular locus consistent with the boundary light-cone having full Lorentz invariance in one lower dimension. We however want a boundary light-cone consistent with a Galilean invariant field theory living in two lower dimensions, which the bulk correlators do not  see unless we explicitly break boundary Lorentz invariance. On the contrary the geometries \req{vacmet} will indeed give a singular locus of the correlators which is commensurate with a Galilean light-cone as discussed in ref.\ \cite{Hubeny:2005qu}.

Another point in favour of the original proposal of
refs.\ \cite{Son:2008ye,Balasubramanian:2008dm} is that in the spacetime
\req{vacmet}, the symmetry is broken to Galilean invariance
irrespective of whether $v$ is compact or not. For $\nu =1$,
compactification of $v$ to obtain a DLCQ description doesn't break any
further symmetry, so the period of compactification $\Delta v$ is a
physical parameter, which can be interpreted as the inverse of the
Galilean mass in the non-relativistic CFT. On the other hand, if the
spacetime is pure AdS, the symmetry is only broken to
the Galilean invariance by compactification in $v$. Prior to
compactification, we have boost invariance $u \to \lambda\, u$ and
$v\to \lambda^{-1} \,v $ in addition to the scaling symmetry
\req{galscaling}. This broken boost invariance can be used to relate
different values of $\Delta v$, making the compactification radius an
unphysical parameter. To be more explicit, we rewrite
\eqref{vacmet} in a more general form, 
\begin{equation}
ds^2 = r^2\, \left(-2 \, du\, dv - \beta^2 r^{2 \nu}\, du^2 + d{\bf x}^2 \right) + \frac{dr^2}{r^2}.
\label{vacmet2}
\end{equation}	
Here we have introduced an additional parameter $\beta$: we can set
$\beta =1$ by a boost $u \to \beta \,u$, $v \to \beta^{-1} v$. If $v$ is
compact, the combination $\beta/ \Delta v$ is invariant under this
boost transformation. In the approach of
refs.\ \cite{Son:2008ye,Balasubramanian:2008dm}, the boost is used to set
$\beta = 1$, and the invariant quantity $\beta/ \Delta v$ is then
interpreted as the Galilean mass. We can now recognize the pure AdS
duality of refs.\ \cite{Goldberger:2008vg,Barbon:2008bg} as the special case
in which we set $\beta =0$, so the boost invariant is zero, and the
coordinate period $\Delta v$ is not a physical parameter. From the
non-relativistic CFT point of view, $\beta=0$ is a limit in which the
Galilean mass vanishes.  This clarifies 
the observation in ref.\ \cite{Goldberger:2008vg} that the Galilean mass does not enter into the formula for operator dimensions in this simplified case.  Pure AdS as a holographic dual is a degenerate special
case of that of \eqref{vacmet}.

We will now proceed to embed \req{vacmet} into string theory and
realize a class of non-relativistic CFTs using conventional D-brane
physics.

\section{Embedding in string theory}
\label{solgen}

The geometry \eqref{vacmet} can be consistently embedded in a solution
to string theory. Indeed, geometries of this type have previously been
studied, in investigations of the application of solution generating
transformations to construct geometries corresponding to twisted
versions of the D3-brane worldvolume theory \cite{Alishahiha:2000pu,
Hubeny:2005qu}. In this section, we first review this solution
generating transformation, and use it to construct a string theory
solution which reduces to \eqref{vacmet} in five dimensions. We then
apply the same transformation to obtain a non-extremal generalization,
and construct a five-dimensional theory for which the non-extremal
geometry is a solution.

\subsection{Generating the geometry dual to the vacuum state}
\label{vacgeom}

To begin with, consider the geometry of \AdS{5}$\times \Sp^5$ in
Poincar\'e coordinates, which is the near-horizon geometry of
D3-branes in flat space:
\begin{eqnarray}
ds^2 &=& r^2\, \left( -dt^2 + d{\bf x}^2 + dy^2 \right) + \frac{dr^2}{r^2} +  (d\psi + A)^2 + d\Sigma_4^2, \nonumber \\
F_{(5)} &=&  dC_{(4)} = 2\, (1+\star) \,d\psi \wedge J \wedge J,
\label{adsmet}
\end{eqnarray}	
where we have written the metric on the unit $\Sp^5$ as a fibration
over a ${\bf CP}^{2}$ base and now ${\bf x} = \{x_1, x_2\}$. The five-form
is given explicitly in terms of the volume form of $\Sp^5$, which has
been decomposed into quantities related to the fibration. $J$ is the
K\"ahler form on ${\bf CP}^2$ and $A$ is the associated potential. Our
conventions are
\begin{equation}
dA = 2\,J \ , \qquad {\rm Vol} \left({\bf CP}^2\right) = \frac{1}{2}\, J\wedge J.
\end{equation}	

We apply a Null Melvin Twist to this geometry, as described in
\cite{Gimon:2003xk}; the idea is to generate light-like NS-NS flux by
a series of boosts and twisted T-dualities. Algorithmically we proceed
as follows:\footnote{The D3-brane geometry above has a full $SO(1,1)$
symmetry in the $(t,y)$ plane which renders the first step
inconsequential here, but it will be meaningful for the non-extremal
solution which follows.}
\begin{enumerate}
\item Pick a translationally invariant direction (say $y$) and boost by amount $\gamma$ along $y$.
\item T-dualize along $y$.
\item Twist some one-form $\sigma$: $\sigma \to \sigma + \alpha \, dy$.
\item T-dualize along $y$ again.
\item Boost by $-\gamma$ along $y$. 
\item Scale the boost and twist: $\gamma \to \infty$ and $\alpha \to 0$, keeping
\begin{equation}
\beta  = \frac{1}{2}\,\alpha \, e^\gamma = {\rm fixed.}
\label{nullmscale}
\end{equation}
\end{enumerate}

The only data needed to describe the construction is the choice of the
one-form $\sigma$. We can choose $\sigma$ to be along the world-volume
directions (linear combination of $dx_1$ and $dx_2$) or transverse to
the D-brane. The former leads to turning on constant electric and
magnetic fields on the D-brane world-volume leading to a light-like
non-commutative field theory \cite{Alishahiha:2000pu,
Hubeny:2005qu}. Due to the presence of world-volume fluxes these
geometries break the rotational invariance in the ${\bf x}$ directions
and they also give rise to geometries \req{vacmet} with $\nu \neq 1$ ;
we will not consider them in detail in the main text of the paper, but
discuss aspects of these geometries in \App{fluxbgs}.

Twisting along the $R$-symmetry direction is more interesting. A natural choice is  to take the one-form $\sigma$ to be along the fiber direction: $\sigma = d\psi$. The Null Melvin Twist leads to the geometry \cite{Hubeny:2005qu}
\begin{eqnarray}
ds^2 &=& r^2 \, \left( -2\, du \,dv -  r^2\, du^2 +  d{\bf x}^2\right) + \frac{dr^2}{r^2}  +  (d\psi + A)^2 + d\Sigma_4^2, \nonumber \\
F_{(5)} &=& 2 \, (1+\star) \, d\psi \wedge J \wedge J , \nonumber \\
B_{(2)} & = & r^2 \, du \wedge (d\psi+ A),
\label{steps56}
\end{eqnarray}	
where the light-cone coordinates are
\begin{equation}
u = \beta \, (t+y) \ , \qquad v = \frac{1}{2\,\beta} \, (t-y). 
\label{rescaleuv}
\end{equation}	
Note that our boosted $uv$ coordinate frame scales $\beta$ out
not only from the metric but also from the field strengths. 
The five-dimensional part of this metric is precisely the geometry
\req{vacmet}, with $\nu =1$ and $d=2$. This geometry will correspond
to the vacuum state of the dual non-relativistic field theory.

The Null Melvin Twist construction makes the interpretation of the
dual field theory clean: it is nothing but $\CN = 4$ Super Yang-Mills
twisted by an R-charge. The $U(1)$ isometry generating the R-charge is
generated in the spacetime by $\frac{\partial}{\partial
\psi}$. This twist breaks the $SU(4)$ symmetry of $\CN =4$ down
to an $SU(3) \times U(1)$ (the isometry group of ${\bf CP}^2$) through the non-vanishing  NS-NS potential $B_{(2)}$ (the metric \req{steps56} of course enjoys full $SU(4)$ invariance).

\subsection{The non-extremal solutions}
\label{nonextgeom}

As we have generated \req{steps56} by a solution generating technique,
we can just as well generate the non-extremal version of the
solution. To do so, rather than starting with the near horizon geometry
of extremal D3-branes, we start with non-extremal D3-branes and repeat
the Null Melvin Twist.  Consider then the planar Schwarzschild-AdS
black hole (times $\Sp^5$, with the geometry supported by the
five-form flux $F_{(5)}$)
\begin{equation}
ds^2 = r^2\left(-f(r) \, dt^2 + dy^2 + d{\bf x}^2 \right) + \frac{1}{r^2} \left(\frac{dr^2}{f(r)} + r^2 \, d\Omega_5^2\right)  ,
\label{nonextD3}
\end{equation}	
where as before we will write the $\Sp^5$ as a $\Sp^1$ fibration over
${\bf CP}^2$. The Null Melvin Twist leads to the string frame metric
\cite{Nayak:2004rc}:
\begin{eqnarray}
ds_{str}^2 &=&  r^2\, \left(- \frac{\beta^2 \, r^2 \, f(r)}{k(r)} \, (dt+dy)^2 -\frac{f(r)}{k(r)}\, dt^2 + \frac{dy^2}{k(r)}  + d{\bf x}^2 \right)+ \frac{dr^2}{r^2\, f(r)} +\frac{(d\psi+A)^2}{k(r)}+ d\Sigma_4^2, \nonumber \\
e^\varphi &=& \frac{1}{\sqrt{k(r)}} ,\nonumber \\
F_{(5)} &=&  dC_{(4)} = 2\, (1+\star) \,d\psi \wedge J \wedge J, \nonumber \\
B_{(2)} & =& \frac{r^2\, \beta}{k(r)}\, \left( f(r)\, dt + dy\right) \wedge (d\psi+ A),
\label{nonextsol}
\end{eqnarray}	
with
\begin{equation}
f(r) = 1- \frac{r_+^4}{r^4} \ , \qquad k(r) = 1 +\beta^2 \,r^2 \,(1-f(r))  = 1 + \frac{\beta^2 \, r_+^4}{r^2}.
\label{fkdefs}
\end{equation}	
The solution has a horizon at $r= r_+$. Note that the parameter
$\beta$ appearing in this metric is an independent physical parameter;
in the extremal case, we could set it to one by boosting in the 
$ty$ plane, but non-extremality has broken this boost symmetry. The
remainder of the paper will be devoted to an exploration of the
physics of this non-extremal solution. First, in the next section, we
construct an appropriate five-dimensional Lagrangian which has
\eqref{nonextsol} as a solution.

\subsection{Five dimensional effective Lagrangian}
\label{relson}

The solutions we have discussed above \req{steps56} and
\req{nonextsol} satisfy the 10-dimensional Type IIB equations of
motion. In \cite{Son:2008ye}, the vacuum geometry \req{vacmet} was
considered as a solution to Einstein-Proca theory with negative
cosmological constant, which has the action
\begin{equation}
\SS_{EP} = \int d^{d+2}x\, dr \, \sqrt{-g} \, \left( R - 2\, \Lambda -\frac{1}{4}\, F_{\mu \nu}\, F^{\mu \nu} - \frac{1}{2}\, m^2 \, A^\mu \, A_\mu \right) \ ,
\label{sonlag}
\end{equation}	
with $F_{\mu \nu} = 2\, \nabla_{[\mu} A_{\nu]}$. The metric
\req{vacmet} with $A^v = 1$ satisfies the field equations for the
choice
\begin{equation}
\Lambda = -\frac{1}{2}\, (d+1)(d+2)\ , \qquad m^2 = 2\, (d +2) \ .
\label{sonLm}
\end{equation}	

We would now like to understand the relation between this
phenomenological Lagrangian and the ten-dimensional IIB theory.
Starting from Type IIB supergravity, we can KK reduce the solution
\req{steps56} on the $\Sp^5$ (which is undeformed). The reduction of
the metric is straightforward, and gives \eqref{vacmet} in five
dimensions. The NS-NS two-form, however, depends on the $\Sp^5$
coordinates. In a linear analysis \cite{Kim:1985ez}, such a mode of
the two-form produces a massive vector transforming in the ${\bf 15}$
of $SO(6)$: in AdS units (set here to 1) its mass is $m^2 =8$. This is
precisely the value of the mass required according to \req{sonLm}
(with $\Lambda = -6$ as necessary to get AdS radius equal to
1). 

From the CFT point of view, this massive vector field in the bulk corresponds to a dimension 5 operator in $\CN = 4$ SYM.  The
twist by R-symmetry is by an irrelevant operator  of dimension $5$ transforming in the antisymmetric tensor representation of $SU(4)$. The operator in question \cite{Alishahiha:2003ru} is 
$\CO_\mu^{IJ} = {\rm Tr} \left( {F_\mu}^\nu\, \Phi^{[I}\, D_\nu \Phi^{J]} + \sum\limits_K \, D_\mu\, \Phi^K\, \Phi^{[K}\Phi^I \Phi^{J]} \right) + {\rm fermions}$, where $\Phi^I$ are the adjoint scalars of $\CN = 4 $ SYM transforming in the vector ${\bf 6}$ of $SU(4)$ and $F_{\mu \nu}$ is the gauge field strength. The  Lorentz symmetry is broken by adding $\CO^{IJ}_u$ to the field theory Lagrangian. This  field theory realization makes it clear that the massive vector
used in the construction of \cite{Son:2008ye} oxidises to NS-NS flux
in ten dimensions.

It is, however, important to note that this massive vector is not part
of gauged supergravity in five dimensions. Thus, it is not obvious
that \eqref{sonlag} is a consistent truncation of the ten-dimensional
theory. That is, while we have found an embedding of \eqref{vacmet} in
the ten-dimensional geometry \eqref{steps56}, we have no guarantee
that solutions of \eqref{sonlag} can in general be oxidised to
solutions of the ten-dimensional IIB equations of motion.

If we perform the same Kaluza-Klein reduction for the non-extremal
solution \eqref{nonextsol}, we obtain 
\begin{eqnarray}
ds_E^2 &=&  r^2\,k(r)^{-\frac{2}{3}} \,\left( -\beta^2 \,r^2 \,f(r) \,(dt+dy)^2 - f \,dt^2 + dy^2 + k \,d {\bf x}^2 \right) 
+  k(r)^{\frac{1}{3}}\, \frac{dr^2}{r^2\, f(r)} \ , \nonumber \\
&=& r^2\, k(r)^{-\frac{2}{3}}\left(\left[\frac{1-f(r)}{4\beta^2} -
    r^2\,f(r)\right] \, du^2 + \frac{\beta^2 r_+^4}{r^4} \, dv^2 - \left[1+f(r)\right]\,du\,dv \right) \nonumber \\
&& \quad+\;\;  k(r)^{\frac{1}{3}}\, \left(r^2 d {\bf x}^2 +  \frac{dr^2}{r^2\, f(r)} \right),
\label{5dbh}
\end{eqnarray}
where we have introduced the light-cone coordinates \eqref{rescaleuv}
in the second line for future convenience, with the massive vector and
scalar
\begin{eqnarray}
A &=& \frac{r^2 \beta}{k(r)} (f(r) \, dt + dy) =\frac{r^2 }{k(r)} \, \left( \frac{1+f(r)}{2}\, du - \frac{\beta^2 r_+^4}{r^4}\, dv\right), \nonumber \\
 e^\phi &=& \frac{1}{\sqrt{k(r)}} \ ,
\label{5doth}
\end{eqnarray}
where $f(r)$ and $k(r)$ are given in \req{fkdefs}. Note that in these
light-cone coordinates, the solution asymptotically approaches the
extremal solution \eqref{vacmet}, but $\beta$ remains a physical
parameter, as the full metric depends on $\beta$. We will henceforth
work with the solution \req{5dbh}.

This black hole solution is not a solution of \eqref{sonlag}, as it
contains a scalar field. However, it is a solution of the equations of
motion from the effective 5 dimensional action
\begin{equation}
\SS = \frac{1}{16 \pi \,G_5}\,\int d^5 x \sqrt{-g} \left(R - \frac{4}{3} (\partial_\mu \phi) (\partial^\mu \phi) - \frac{1}{4} e^{-8 \phi / 3} F_{\mu\nu} F^{\mu\nu} 
- 4\, A_\mu A^\mu - V(\phi) \right) ,
\label{5deff}
\end{equation}
where the scalar potential is
\begin{equation}
V(\phi) = 4 \,e^{2 \phi/3} (e^{2 \phi} - 4),
\label{Vpot}
\end{equation}	
The scalar here appears from two sources: (i) the black hole geometry
involves a non-vanishing dilaton and (ii) the twist now causes the
fibration over ${\bf CP}^2$ to be squashed. Squashing is a common feature
of solutions generated by the Null Melvin Twist \cite{Gimon:2003xk}
and intuitively can be ascribed to the distortion of the asymptotics
of the spacetime.

To summarize, it is easy to embed geometries of the form \req{vacmet}
into string theory. The dimensionally reduced descriptions appear to
involve exotic matter like Proca fields as in \req{5deff}, but their
stringy origin is simply in terms of conventional supergravity
$p$-forms.

Again, we do not have any argument that the five-dimensional action
\eqref{5deff} describes a consistent truncation of the full
ten-dimensional theory, and indeed, one might in general expect the
modes transforming non-trivially under $SO(6)$ which are turned on in
our ansatz to couple to other Kaluza-Klein harmonics which we have
neglected. However,  by construction we know this particular five-dimensional solution (\ref{5dbh}) uplifts to type IIB supergravity, and from now on, we will work with the
five-dimensional action \eqref{5deff}.  In Appendix \ref{fluxbgs}, we discuss a
different way of realizing geometries like \eqref{vacmet}, using
$p$-form fluxes which break the rotational invariance in the spatial
directions. This alternative approach can be easily embedded in
five-dimensional gauged supergravity, using a Null Melvin Twist along
the worldvolume directions.

\section{Thermodynamics from gravity}
\label{therdof}

We are interested in understanding the thermodynamics of the black
hole solution \req{5dbh}. The simplest thing to understand is the
entropy of the black hole. The geometry \req{5dbh} has a horizon
at $r= r_+$ and one can compute the area of this horizon. In fact,
since we generated the solution by a series of boosts and dualities,
it turns out that the horizon area is independent of $\beta$ \cite{Gimon:2003xk}. We obtain thus
\begin{equation}
S = \frac{1}{4\,G_5} \,r_+^3 \,{\rm Vol}(horizon) =
\frac{1}{4\,G_5} \, r_+^3 \,\Delta y \,\Delta x_1 \,\Delta x_2.
\label{entropy}
\end{equation}	

We can also compute the Hawking temperature of the black hole, which
is most simply done by computing the surface gravity of the
horizon. Given the null generator of the horizon, $\xi^a$, the surface
gravity $\kappa$ is defined as
\begin{equation} 
\kappa^2 = - {1 \over 2  } \, \left( \nabla^a \xi^b \right) 
\left(\nabla_a \xi_b \right),  
\label{surfgrav}
\end{equation}
with the evaluation at the location of the horizon implicit. The
Hawking temperature is given in terms of the surface gravity as $ T_H
= {1 \over 2 \, \pi } \, \kappa$.

It is clear from \req{5dbh} that the Killing generator of the event
horizon is proportional to $\frac{\partial}{\partial
    t}$. To fix the constant of proportionality, we require
that the component along the boundary time-translation have
coefficient one. The generator of time translation for our
non-relativistic CFT is $\frac{\partial}{\partial {
      u}}$ in the light cone coordinates \eqref{rescaleuv},
so this requirement fixes the normalization of the Killing generator
of the horizon as
\begin{equation}
\xi = \frac{1}{\beta}\,\frac{\partial}{\partial t} =\frac{\partial}{\partial u} + \frac{1}{2 \beta^2} \,\frac{\partial}{\partial v}.
\label{normKgen}
\end{equation}	
With this definition we find that 
\begin{equation}
  T = \frac{r_+}{\pi \,\beta}.
\label{bhtemp}
\end{equation}	
The Killing generator of the event horizon \req{normKgen} not only has
components along the boundary time translation direction $u$, but also
along the light-like direction $v$.  From the gravitational
perspective it therefore appears that we are dealing with a system
where we have a chemical potential for $v$-translations
\begin{equation}
\mu = \frac{1}{2 \beta^2} \ .
\label{mudef}
\end{equation}	
The conserved charge conjugate to this chemical potential is just
$\frac{\partial}{\partial v}$ momentum. We therefore
claim that the black hole solution \req{5dbh} corresponds to
thermodynamics described by the density matrix for a grand canonical
ensemble
\begin{equation}
\rho = \exp{\left(-\frac{\hat{H}}{T} - \frac{\mu \,\hat{P}_v}{T}\right) }.
\label{densitymat}
\end{equation}	
In the dual field theory, the operator generating translations in $u$
is the Hamiltonian $\hat{H}$, while the $v$ momentum corresponds to
the integer quantized particle number $\hat N$, $\hat P_v = 2\pi \hat N/\Delta v$. Note
that the operator $\hat{N}$ commutes with the Hamiltonian ${\hat
  H}$. In fact the only place it shows up in the Galilean conformal
algebra \cite{Nishida:2007pj} is in the commutator of spatial momentum
and Galilean boosts, $[\hat P^i , \hat K^j ] = -i \delta^{ij} \, \hat N$, as it
commutes with all the other generators of the Schr\"odinger
algebra. The Gibbs potential of this ensemble $\CQ(T,\mu,V)$ can be
calculated from the partition function
\begin{equation}
\Xi(T, \mu) = {\rm Tr}(\rho) \ , \qquad   {\mathcal Q}(T, \mu ,V) =
-T\, \log \, \Xi(T,\mu).
\end{equation}	
We can then pass to a more conventional ensemble with fixed particle
number $N$ by a Legendre transformation leading to a free energy
$F(T,N,V)$ in the canonical ensemble
\begin{equation}
F(T,N,V) = {\mathcal Q}(T,\mu,V) - \mu\, N.
\end{equation}	

The non-relativistic theories we are dealing with are realized as a
deformed version of a relativistic quantum field theory. Hence it is
not surprising that the natural ensemble is one where the particle
number is allowed to fluctuate.  Another argument for the naturalness
of the grand canonical ensemble comes from the
geometry. The Galilean theories we constructed are
embedded into a higher dimensional Poincar\'e invariant theory (in the
present case $\CN =4$ SYM), in which we turn on some background fields
to break Poincar\'e invariance. We can reduce the theory on the
light-cone to obtain a Galilean CFT, but in this procedure there is no
canonical choice of the $P_v$ eigenvalue $2 \pi N/\Delta v$. It therefore
seems natural to sample over the space of eigenvalues weighted by some
parameter $\mu$. 

\section{Asymptotics and action}
\label{asymcons}

We now turn to the calculation of the Gibbs potential $\CQ(T,\mu,V)$
in a saddle-point approximation. This potential can be obtained from the
on-shell action of an analytically continued version of the black hole
solution. The geometry \req{5dbh} does not have a real Euclidean
section because of the presence of non-zero chemical potentials, but
we can still use the analytically continued metric in a saddle-point
approximation: as argued in  \cite{Brown:1990fk} (in the context of the
Kerr-Newman solutions), the appropriate saddle point for the
thermodynamic partition function is obtained by analytic continuation
of the time coordinate, even in cases where the resulting metric is complex. 
The action evaluated on this analytically continued solution
is always real, so it can be used as a saddle-point approximation to the
thermodynamic potential. 

We would also like to be able to derive a boundary stress tensor using
an extension of the Brown-York type analysis used in the context of
AdS/CFT \cite{Henningson:1998gx,Balasubramanian:1999re}. It is not
clear that this technique can be straightforwardly applied to our
solution, because of the inhomogeneity in the asymptotic falloff
conditions for different components of the metric, which implies that
we do not have a regular conformal structure on the
boundary.\footnote{From the discussion of the causal properties in
  \sec{rev} it follows that we also cannot define a causal boundary
  for \req{vacmet}.} However, a first step in such a
calculation is the determination of the necessary boundary terms
required to obtain a well-behaved action. A well defined action should
have vanishing variation on-shell in a classical phase space that
encompasses the solutions we are interested in.

We will start by analyzing the asymptotic fall-off conditions for the
black hole spacetime. We then proceed to construct an action that is
stationary with respect to an appropriate class of variations. Our
main result is summarized in \req{finaction}, which we then use to
extract the Gibbs potential $\CQ(T,\mu,V)$ from a saddle point
evaluation.

\subsection{Naive asymptotics from metric expansion }
\label{asymf}

To understand the asymptotics, let us focus on the five dimensional
solution \req{5dbh} in the light-cone coordinates \req{rescaleuv}. We
find
\begin{equation}
\begin{aligned}
g_{{  u}{  u}} &=   - r^4 + \frac{2}{3}\, \q\, r^2 + \ord{1} , & \qquad 
g_{{  u}{  v}} &=  -r^2  + \frac{2}{3} \, \q +\ord{r^{-2}} ,\\
g_{{  v}{  v}} &=  \frac{\q}{r^2} +\ord{r^{-4}}, &\qquad 
g_{{\bf x}{\bf x}} & = r^2 + \frac{1}{3}\,\q + \ord{r^{-2}},  \\
g_{rr} &= \frac{1}{r^2} + \frac{1}{3}\, \frac{\q}{r^4} + \ord{r^{-6}}, &\qquad  &
\end{aligned}
\label{asymexp}
\end{equation}	
and for the inverse metric,
\begin{equation}
\begin{aligned}
g^{{  u}{  u}} &=  -\frac{\q}{r^6} +\ord{r^{-8}}, &\qquad
g^{{  u}{  v}} &=  -\frac{1}{r^2}  + \frac{1}{3} \frac{\q}{r^4} +\ord{r^{-6}} ,\\
 g^{{  v}{  v}} &=   1 - \frac{1}{3}\frac{\q}{r^2} + \ord{r^{-4}} , & \qquad 
g^{{\bf x}{\bf x}} & = \frac{1}{r^2} - \frac{1}{3} \frac{\q}{r^4} + \ord{r^{-6}},  \\
g^{rr} &= r^2 - \frac{1}{3} \q  + \ord{r^{-2}}, &\qquad  &
\end{aligned}
\label{asymexp2}
\end{equation}	
where for ease of notation, we introduce $ \q \equiv \beta^2 \,
r_+^4$, which encodes the leading deformation of the vacuum spacetime
\req{vacmet}. For the matter fields it follows from \req{5doth} that
\begin{equation}
\begin{aligned}
A_{  u} & = r^2 - \q  + \ord{r^{-2}} , & \qquad
A_{  v} & = -\frac{\q}{r^2} + \ord{r^{-4}} ,\\
\phi\; & = -\frac{\q}{2 \,r^2} + \ord{r^{-4}}. & \qquad &
\label{Aphiasym}
\end{aligned}
\end{equation}	
It is worth remarking that these fall-off conditions are a result of a
coupling between the linearized fluctuations of the fields about the vacuum
background \req{steps56}; the fluctuations of the gravitational, vector and scalar degrees of
freedom do not decouple in the five-dimensional
effective action \req{5deff}.

\subsection{A stationary action}
\label{staction}

We want to evaluate the on-shell action for the solution
\eqref{5dbh}. To get a well-behaved action, we need to supplement the
bulk action \eqref{5deff} with boundary terms to satisfy the condition
that $\delta \SS = 0$ on-shell, for variations satisfying suitable
falloff conditions. In this subsection, we construct an action
satisfying this condition for a very restricted set of variations ---
the minimal set including variations along the family of solutions
we're interested in.

We will construct an action principle by adding local covariant
counterterms to the bulk action plus a Gibbons-Hawking boundary term, as
in asymptotically AdS spacetimes
\cite{Henningson:1998gx,Balasubramanian:1999re}. If we add the most
general combination of local counterterms which can make non-zero
contributions to the on-shell action, we have
\begin{eqnarray}\label{adsact}
  \SS &=& \frac{1}{16\pi\, G_5}\,\int d^5 x \,\sqrt{-g} \left(R - \frac{4}{3} (\partial_\mu \phi) (\partial^\mu \phi) - \frac{1}{4} \,e^{-8 \phi / 3}\, F_{\mu\nu} F^{\mu\nu} 
    - 4 \,A_\mu A^\mu - V(\phi) \right) \nonumber \\
  &&+ \frac{1}{16\pi  \,G_5} \int d^4 \xi\,
  \sqrt{-h} \,\left(2K - 2\,c_0 + c_1 \,\phi + c_2\,
    \phi^2  \right.
   \nonumber  \\
&& \qquad \qquad \qquad \qquad  \qquad \qquad \qquad   \left. + c_3 \,A_\alpha A^\alpha + c_4\, A_\alpha A^\alpha \phi + c_5 \, (A_\alpha A^\alpha )^2 \right)\ .
\label{startS}
\end{eqnarray}
Here $\xi^\alpha$ are boundary coordinates, $h_{\alpha \beta}$ is the
induced metric on the boundary, and $K = K_{\alpha \beta} h^{\alpha
  \beta}$, with $K_{\alpha \beta}$ the extrinsic curvature. We want to
fix the coefficients $c_0, c_1, c_2, c_3, c_4$ by imposing $\delta \SS=0$. In
general, the variation of this action is
\begin{eqnarray}
\label{genvar}
\delta \SS &=& \frac{1}{16\pi \,G_5} \int d^4 \xi \,\sqrt{-h} \, \left[ 
\left(
\pi_{\alpha \beta} + 
\left( 
c_0 - \frac{1}{2} \phi \, c_1 - \frac{1}{2} \phi^2 c_2  
\right) h_{\alpha \beta}
\right. \right. \nonumber \\
&& 
\qquad \qquad \qquad \qquad +\left(
A_\alpha A_\beta-
\frac{1}{2} \,A_\gamma A^\gamma \,h_{\alpha \beta} 
\right) (c_3 + \phi \, c_4)
 \nonumber \\
&&
\qquad \qquad \qquad \qquad
 \left.
+\left( 2 A_\alpha A_\beta - \frac{1}{2} A_\gamma A^\gamma h_{\alpha \beta}
\right)c_5  \left(A_\delta A^\delta \right)
\right) \,\delta h^{\alpha \beta} 
 \nonumber \\
 && 
 \qquad \qquad \qquad \qquad
+ \Bigl(
- n^\mu F_{\mu \alpha} e^{-8\phi/3}
+2( c_3 +  \phi \, c_4 + 2 c_5 (A_\gamma A^\gamma)^2 ) A_\alpha
\Bigr) \delta A^\alpha 
\nonumber \\
&&
\qquad \qquad \qquad \qquad
\left.
+ \left( 
 - \frac{8}{3} n^\mu \partial_\mu \phi + \left(c_1 + 2 \phi \, c_2 + A_\alpha A^\alpha \, c_4 \right)
\right)
\delta \phi 
\right]
\end{eqnarray}
where $n^\mu$ is the unit normal to the boundary, and $\pi_{\alpha
  \beta} = K_{\alpha \beta} - h_{\alpha \beta} K$. 

We need to establish an appropriate class of variations. To do so, we
need to define the phase space of solutions: that
is, we need to specify the asymptotic boundary conditions on the
fields. We choose our asymptotic boundary conditions to require that
the leading falloff agrees with the $\q$ independent terms in
eqs.\ (\ref{asymexp}, \ref{asymexp2}, \ref{Aphiasym}). This requirement implies that
no variation of these terms is allowed; these terms are non-dynamical.
We further require that the
subleading terms in the asymptotic falloff be related as in
eqs.\ (\ref{asymexp}, \ref{asymexp2}, \ref{Aphiasym}). This restrictive
but permissible choice of boundary 
conditions by construction admits our black hole
as an allowed solution. More precisely, the leading
non-zero variation is required to be of the form 
\be 
\delta h_{ab} =
\frac{d h_{ab}}{ d\q }\delta a \;\; , \; \; \delta A_a = \frac{d
  A_a }{d\q} \delta a \; \; , \; \; \delta \phi = \frac{ d \phi
}{d\q} \delta a \ ,
\label{fulldef}
\ee 
rather than allowing independent variations of the different
fields at this order.  We will denote this variation of the
fields collectively by $\delta \psi_d$. 
In addition to this variation, we allow arbitrary
variations $\delta \psi_f$ where the variations of $\delta h_{ab}$,
$\delta A_a$ and $\delta \phi$ are treated as independent and fall off
at least one power of $r^2$ faster than in \eqref{fulldef}.  Given
that these $\delta \psi_f$ are independent, instead of considering the
full variation 
(\ref{fulldef}), we are free to take a linear combination of $\delta
\psi_d$ and the $\delta \psi_f$ and replace $\delta \psi_d$ with its
leading order behavior:
\begin{equation}
\begin{aligned}
\delta \psi_d \; : \; \; \; &&
  \delta h^{uu} &= -\frac{1}{r^6} \delta a, &\quad \delta h^{uv} &=
  \frac{1}{3 r^4} \delta a, &\quad \delta h^{ v
    v} &= - \frac{1}{3r^2} \delta a,\\
&&  \delta \phi &= - \frac{1}{2 r^2} \delta a, &\quad \delta A_{ u} &= -
  \delta a, &\quad \delta A_{ v} &= - \frac{1}{r^2} \delta a.
\end{aligned}
\label{restvarA}
\end{equation}

Substituting the asymptotic fields (\ref{asymexp}, \ref{asymexp2},
\ref{Aphiasym}) into the general variation of the action
\eqref{genvar}, we find that the terms in $\delta \SS$ which are
independent of $\q$ are 
\begin{eqnarray}
\label{deltaSrsq}
\delta \SS|_{\q=0} &=& \frac{1}{16\pi \,G_5} \int d^4 \xi \,\sqrt{-h} \, \left\{
  r^4 \,(2+c_3-c_0) \,\delta h^{uu} -2\,r^2 \,(c_0-3) \,\delta h^{uv} \right. \nonumber   \\ &&
 \left. + \,r^2\,  (c_0-3) \left(\delta h^{x_1 x_1} + \delta h^{x_2 x_2}\right) + c_1 \,\delta \phi + 2\,r^2 \,(c_3-1) \delta
  A^u \right\}.
\end{eqnarray}
This part of the variation of the action will receive contributions
that diverge like $r^2$ from $\delta \psi_d$, and finite contributions
from $\delta \psi_f$. Since the variations of the different fields in
$\delta \psi_f$ are independent, we need to set $c_0=3$, $c_1 =
0$, and $c_3 = 1$ to cancel these variations. 

We are then left with contributions to $\delta \SS$ which go like
$\q$, 
\begin{eqnarray}
\label{deltaSfin}
\delta \SS &=& \frac{\q}{16\pi \,G_5} \int d^4 \xi \,\sqrt{-h} \,
\left[-\left(\frac{13}{6} + \frac{c_4}{2} - 2 c_5\right) \,r^2 \,\delta h^{  u   u}
    - 4 \,\delta h^{  u   v} - \frac{2}{r^2} \,\delta h^{  v   v} \right.  \nonumber\\
    && \left. +
    \left( c_4 - c_2 -\frac{8}{3} \right)  \frac{\delta \phi}{r^2}  -
    \left(  \frac{13}{3}+c_4 - 4 c_5 \right) \,\delta A^{  u} -
    \frac{4}{r^2}\, \delta A^{v}
  \right].
\end{eqnarray}
These terms will receive finite contributions from the variations
$\delta \psi_d$, and vanishing contributions from the variations
$\delta \psi_f$.  Using (\ref{restvarA}), the total variation of the
action \eqref{deltaSfin} becomes
\begin{equation}
\delta \SS = \frac{\q}{16\pi \,G_5} \int d^4 \xi\,
\sqrt{-h} \;\frac{c_2  -2 c_4 +4 c_5 - 3}{2 \,r^4}\, \delta a \ ,
\end{equation}
which will have the desired vanishing value provided $c_2 - 2\,c _4 +4\, c_5= 3$.

To summarize, an action which satisfies $\delta \SS=0$ for the restricted class
of variations whose leading behaviour is given by \req{restvarA} is
\begin{eqnarray}
\SS &=& \frac{1}{16\pi\, G_5}\,\int d^5 x \sqrt{-g} \left(R - \frac{4}{3} (\partial_\mu \phi) (\partial^\mu \phi) - \frac{1}{4} e^{-8 \phi / 3} F_{\mu\nu} F^{\mu\nu} 
- 4 \,A_\mu A^\mu - V(\phi) \right) \nonumber \\ &&+ \frac{1}{16\pi G_5} \int d^4 \xi \,\sqrt{-h}\, \left(2\, K - 6 
+ A_\mu A^\mu + c_4 \,A_\mu A^\mu \phi + c_5 \left( A_\mu A^\mu\right)^2 \right. \nonumber \\
&&
\left. \qquad \qquad \qquad \qquad \qquad +(2 \,c_4 - 4\, c_5 +3) \phi^2  \right)\ 
\label{finaction}
\end{eqnarray}
for some arbitrary constants $c_4$ and $c_5$. 

\subsection{Euclidean action for the black hole}

Given the action \req{finaction} we can compute its value
on the black hole solution \req{5dbh}, \req{5doth}. We find that the
on-shell value of the action is rather simple,
\begin{equation}
\SS =  \frac{1}{16\pi\, G_5} \int d^4 \xi\; r_+^4\ .
\label{sonshell}
\end{equation}
In fact, it is identical to the on-shell action of the
Schwarzschild-\AdS{5} black hole \req{nonextD3}! Note that the
dependence on $c_4$ and $c_5$ cancels out of the on-shell action, as does the
dependence on $\q$. 

Let's now use this action to compute the thermodynamics of the black
hole in the Euclidean approach. We will assume we compactify $v$ with
period $\Delta v$ and the spatial directions with a volume $V$, and
obtain a ``Euclidean'' solution by analytically continuing $t \to i
\tau$.  Smoothness of this analytically continued solution then forces
the Euclidean time to have period $\Delta \tau = \pi/r_+$, consistent with our identification
of the temperature \req{bhtemp}. Thus, the full Euclidean action is
\begin{equation}
\label{eact}
I =  - \frac{\beta}{16\,G_5} \, r_+^3 \, V\, \Delta   v \ ,
\end{equation}
where we have used the fact that $\beta \Delta v = \Delta y$ at fixed $u$.

We want to interpret this action as the saddle-point
approximation to the grand canonical partition function,
\begin{equation}
\Xi(T,\mu) = e^{-\CQ(T, \mu)/T} =  {\rm Tr}\left(\exp \left( - \frac{\hat{H}}{T} - \frac{\mu \, \hat{P}_v}{T}
\right) \right)\approx e^{-I},
\end{equation}
with temperature $T$ and chemical potential $\mu$ given as in
\req{bhtemp} and \req{mudef} respectively. Note that the Euclidean
action \eqref{eact} is always negative, so the black hole solution
makes the dominant contribution to this partition function for any
non-zero temperature.

Thus, we should be able to extract the expected energy and charge as
\begin{equation}
\langle P_v \rangle = -T \,\frac{\partial}{\partial \mu} \ln \Xi(T,\mu) =  T\,
\frac{\partial}{\partial \mu} I, 
\end{equation}
\begin{equation}
\langle E \rangle + \mu \,\langle P_v \rangle = T^2\,
\frac{\partial}{\partial T} \ln \Xi(T,\mu) = - T^2
\frac{\partial}{\partial T} I. 
\end{equation}
Furthermore, using $\ln \Xi = -\CQ/T$, we should have $\CQ/T = (E+\mu \,P_v)/T - S =
I$. So the entropy should be given by
\begin{equation}
S = - \left( T\, \frac{\partial}{\partial T} +1 \right) I. 
\label{Salt}
\end{equation}
The action written in terms of $T$ and $\mu$ is 
\begin{equation}
I  = -  \frac{\pi^3\, T^3}{64 \,G_5\, \mu^2}
V\, \Delta   v \ ,
\end{equation}
which leads to 
\begin{equation}
S =  \frac{\pi^3 \,T^3}{16 \,G_5\,\mu^2} \,V\, \Delta   v\ , 
\end{equation}
which is the same as the result \eqref{entropy} we obtained earlier by direct
calculation.\footnote{%
 That $S$ is given both in terms of the horizon area and by (\ref{Salt})
  is a consistency check of our calculation: in general, by foliating
  the region outside the horizon by surfaces of constant time, we can
  always rewrite the Euclidean action as $I = \frac{1}{T}(E+\mu \,N) -
  S$, which implies the assumed relation between entropy and action.}

We then obtain the conserved charges
\begin{equation}
\langle N \rangle =  \langle P_v \rangle \frac{\Delta v}{2\pi}   =\frac{\pi^2 \,
  T^4}{64\,G_5\, \mu^3} \;V\, \Delta v^2 \ ,
\label{partno}
\end{equation}
and
\begin{equation}
\langle E \rangle = \frac{\pi^3 \,T^4}{64\, G_5\,\mu^2} \; V\, \Delta   v \ .
\label{energy}
\end{equation}
Furthermore, the pressure is given in the grand canonical ensemble directly in terms of the Gibbs potential $\CQ(T,\mu,V)$:
\begin{equation}
P\, V = - \CQ(T,\mu, V) =  \frac{\pi^3 \,T^4}{64\, G_5\,\mu^2} \; V\, \Delta   v \ ,\label{pressure}
\end{equation}	
leading thus to an equation of state
\begin{equation}
P \, V =  E.
\label{eos}
\end{equation}	
A non-relativistic system with Galilean conformal invariance has
different scalings for temporal and spatial directions as given in
\req{galscaling} for $\nu =1$.  This feature leads to an equation of
state $d\, P\, V = 2 \, E$ in $d$-spatial dimensions \cite{Son:2008ye},
which is satisfied by \req{eos}. So indeed, the black hole solution
constructed describes a state in the grand-canonical ensemble at
temperature $T$ and chemical potential $\mu$ for a non-relativistic
conformal field theory. 

The black hole solution \req{5dbh} has a translationally invariant
horizon in the field theory directions ${\bf x}$ and hence in analogy
with $\CN =4$ thermodynamics one expects that it corresponds to the
high temperature phase of the Galilean CFT. Indeed this expectation is
consistent with the fact that our free energy is always negative. In
\sec{discuss} we discuss the possibility of a Hawking-Page like low
temperature phase transition in finite volume.

\section{Shear viscosity of non-relativistic plasmas}
\label{shear}
  
Strongly coupled non-relativistic plasmas that are encountered in cold
atom systems, i.e.\ fermions at unitarity, are believed to behave
as nearly ideal fluids \cite{Gelman:2004fj,Schafer:2007ib}, like
the quark-gluon plasma. Given that we have a holographic dual which
describes the physics of a strongly coupled non-relativistic system, it
is worth inquiring whether the shear viscosity of the
plasma takes the universal value
$\eta/s = 1/4 \pi$ typical of such holographic systems 
\cite{Kovtun:2004de}.\footnote{The bulk viscosity of the non-relativistic
conformal plasmas vanishes due to the scale invariance \cite{Son:2005tj,Nishida:2007pj}.} In fact, given that the field theories we consider are similar to
non-commutative Yang-Mills theories where it is known that $\eta/s =
1/4\pi$ \cite{Landsteiner:2007bd}, it is not surprising that we
recover this same value, as we now show.\footnote{%
 The viscosity result was first presented by Adams at the BIRS
workshop, who emphasized that the off-diagonal metric component
behaves as a minimally coupled scalar and that the stress-tensor was
dual to a mode with zero $v$-momentum.
}
  
To compute the shear viscosity, we will use the
Kubo formula.  Consider the following off-diagonal component of the Fourier transformed, retarded, two point
function of the stress-tensor:\footnote{We will focus below mostly on
  the spatial components of the stress-tensor, which are the only
  tensorial objects in a Galilean field theory. The energy density  $T_{uu}$ and 
  mass current $T_{ui}$ can be incorporated if we work
  with a generalized stress-tensor complex.} 
\begin{equation}
G_{12, 12} (\omega, 0) = -i \int \, du \,   d^2 x\, e^{i \omega\,u } \,
\theta(u) \vev{[T_{x_1 x_2}(u, {\bf x}) , T_{x_1 x_2}(0,{\bf 0})]}.
\end{equation}	
The shear viscosity is given by the zero-frequency limit of this two
point function,
\begin{equation}
\eta = - \lim_{\omega \to 0}\; \frac{1}{\omega}\, {\rm Im}\left(  G_{12, 12} (\omega)\right).
\label{shdef}
\end{equation}	

To compute the shear viscosity, we
use the recipes of refs.\
\cite{Son:2008ye,Balasubramanian:2008dm}
to compute this shear component of the two point
function of the stress-tensor in the black hole background \req{5dbh}. 
Generic linearized fluctuations between the fields in the action
\req{5deff} involve coupling between the gravitational, vector and dilatonic degrees of freedom.  Happily, for the
stress-tensor two point function, the dual field in the bulk is the
metric fluctuation $\delta g_{x_1 x_2}$ which decouples from the rest
of the fluctuations at linear order. In fact, it turns out that
$\delta g_{x_1}^{\ x_2}$ satisfies a massless, minimally coupled
scalar equation in the background \req{5dbh}.

Consider then fluctuations of the bulk metric \req{5dbh} in the
spatial directions of the non-relativistic field theory. To obtain the
shear viscosity we only need to know the zero momentum ${\bf p} = 0$
value of the correlator. Decomposing the fluctuation $\delta
g_{x_1}^{\ x_2}$ into Fourier components and setting ${\bf p} = 0$,
\begin{equation}
\delta g_{x_1}^{\ x_2} \equiv e^{-i \,\omega \, u + i\, n\, v } \,\chi(\omega, r) \ ,
\end{equation}
we have
\begin{equation}
\frac{f(r)}{r^5}\,\frac{d}{dr}\,\left(r^5\, f(r) \, \frac{d\chi}{dr}\right) + V_{eff}(r) \, \chi(\omega, r) = 0 ,
\label{scwave}
\end{equation}
where
\begin{eqnarray}
V_{eff}(r) &=& \frac{1}{r^4}\, \left( (1-f)\, \beta^2 \omega^2 - (1+f)\,
  \omega \, n + \left(\frac{1}{4} \,(1-f) - \beta^2\,r^2\,f \right) \, \frac{n^2}{\beta^2} \right).
\label{sceffV}
\end{eqnarray}	

We have written the above expressions for general values of the
$v$-momentum of the mode, which we call $n$. However, the stress
tensor of the non-relativistic CFT must correspond to the $n=0$ mode
of the bulk metric. This can be seen in two different ways: first, as
we argued previously, $v$-momentum corresponds to particle number in the
non-relativistic CFT, and the stress tensor does not carry particle
number. Second, the conformal dimension of the operators in the
non-relativistic CFT will explicitly contain $n$ dependence; for a
massless minimally coupled bulk scalar field one has
\cite{Son:2008ye,Balasubramanian:2008dm}
\begin{equation}
\Delta = 2 + \sqrt{n^2 + 4}.
\label{}
\end{equation}	
Such a scalar will have the same conformal dimension as the stress
tensor only for $n=0$. 
(The conformal dimension of a Galilean CFT in $d$ spatial dimensions is $d+2$.)

For $n=0$, we find that \req{scwave} simplifies considerably. In fact,
it becomes identical to the wave-equation for a massless scalar field
at zero momentum on the Schwarzschild-\AdS{5} black hole up to a trivial rescaling of the frequency by $\beta$.  To calculate the Green's function, we solve as usual by
demanding ingoing boundary conditions at the horizon, in the hydrodynamic
limit $\omega /r_+ \ll 1$. Defining $\zeta(r)$ such that
\begin{equation}
\chi(\omega, r) \equiv A \, \left(r-r_+\right)^{- \frac{i \, \beta \,\omega}{ 4\, r_+}} \, \zeta(r) \ ,
\label{chi}
\end{equation}
where the constant $A$ is related to the  boundary
value of $\chi(\omega,r)$, $\chi_0(\omega) =  \chi(\omega, \infty)$. Solving for
$\zeta(r)$ perturbatively in $\omega$, we find that
\begin{equation}
\zeta(r) = 1 - \frac{i\, \beta\,\omega}{4 \,r_+} \ln \left( \frac{f}{r-r_+}
\right) + {\mathcal O}((\omega / r_+)^2) \ .
\label{zeta}
\end{equation}
The two-point function is then determined by the boundary term in the
action for $\chi$
\begin{equation}
\SS_\chi = - \lim_{r\to\infty}  \frac{1}{16 \pi \,G_5} \int \frac{d\omega}{2\pi} \, dv \, d^2{\bf x} \; \frac{1}{2} r^5 \,f(r)  \,\chi(-\omega, r)\,\frac{d\chi(\omega, r)}{dr}  + \ldots 
\end{equation}
We follow the recipe presented in ref.\ \cite{Son:2002sd}.  
(See ref.\ \cite{Herzog:2002pc} for a more rigorous treatment.)
Using eqs.\ (\ref{chi}) and (\ref{zeta}), the boundary term can be written 
\begin{equation}
\SS_\chi = \frac{1}{16\pi \, G_5} \int \frac{d\omega}{2\pi} \, dv \, d^2{\bf x} \, \chi_0(-\omega)\left( \frac{i}{2} r_+^3 \beta \omega \right)\chi_0(\omega)
\end{equation}
from which the hydrodynamic retarded Green's function may be extracted:
\begin{equation}
G_{12,12}(\omega) = -\frac{i}{16 \pi\,G_5} \;\beta\, \omega \, r_+^3 \,\Delta v \ .
\end{equation}
The Kubo formula for the viscosity then 
leads to 
\begin{equation}
\eta = \frac{1}{16 \pi\, G_5} \; \beta\, r_+^3\, \Delta v
\label{}
\end{equation}	
which, noting that $\beta \Delta v = \Delta y$ at fixed $u$, gives 
\begin{equation}
\frac{\eta}{s} = \frac{1}{4\,\pi} \ .
\label{visbnd}
\end{equation}	
The result is in large part a consequence of the fact that the zero-frequency
limit of the two-point function has a trivial $\beta$ dependence, as does the
entropy density obtained from \eqref{entropy}. 
 
\section{Discussion}
\label{discuss}

We have discussed aspects of holography for non-relativistic conformal
field theories, concentrating in particular on the nonzero temperature
physics of these systems. We described how non-relativistic field
theories arise naturally in string theory from the world-volume theories on
D-branes.  The specific construction we focused on is DLCQ
quantization of an R-charged twisted D3-brane world volume theory
($\CN =4$ SYM). However, the construction via the Null Melvin Twist
makes it clear that one can generate a whole host of such theories
with and without conformal invariance.

To generate non-relativistic conformal field theories in $d=2$ spatial
dimensions with dual geometries of the form \req{vacmet} with $\nu
=1$, one can start with any $\CN =1$ superconformal field theory with
an AdS dual. The infinite class of $\CN=1$ quiver gauge theories with
\AdS{5}$\times X_5$ duals, where $X_5$ is a Sasaki-Einstein manifold
\cite{Morrison:1998cs}
 (a special case of which is
the Klebanov-Witten conifold theory \cite{Klebanov:1998hh}) provide
excellent starting points for such constructions. In the dual
spacetime one has a $U(1)_R$ realized as an isometry in $X_5$ analogous
to the $\Sp^1$ fibration over ${\bf CP}^2$ used here. The Null Melvin
Twist along this isometry will yield the appropriate pp-wave
geometries, and
what we have discussed in the text
can be applied to Sasaki-Einstein spaces with little or no modification.
It is easy to see that the non-relativistic equation of state
\req{eos} will be respected by these field theories.
Similarly, one could start with relativistic field theories which have
a non-trivial RG flow and construct non-conformal analogs of Galilean
field theories.  All these examples
provide $2$-spatial dimensional non-relativistic field theories.

It should be possible to generate solutions in higher dimensions $d >
2$ using other D-brane world-volume theories, but these would
typically not exhibit Galilean conformal symmetry. To obtain a
Galilean CFT in $d=3$ spatial dimensions, we would need to start with
an asymptotic \AdS{6} geometry, which would correspond to the strongly
coupled fixed point of a five dimensional CFT.  Other generalizations of course include the M-brane world-volume CFTs, wherein the asymptotic \AdS{4} and \AdS{7}
spacetimes can be twisted with the R-symmetries, which is
geometrically achieved by turning on 3-form fluxes in the background.

Even without the precise non-extremal geometries at hand, one can
infer some thermodynamic features of the higher dimensional
relativistic CFTs, if we assume that as in the case studied here, the
gravitational and matter degrees of freedom conspire to give a
Euclidean action which agrees with the action for the untwisted
asymptotically AdS black hole. In $d$-spatial dimensions we would then
have (gathering all the irrelevant numerical coefficients in $\Gamma$ and $\Gamma'$)
\begin{equation}
I =  -\Gamma' \, \beta \, r_+^{d+1} = - \Gamma\, \frac{T^{d+1}}{\mu^{\frac{d}{2} +1}}  \ . 
\label{}
\end{equation}	
The scaling here follows from $T = \frac{(d+2) \, r_+ }{4\, \pi \,
  \beta}$ and we expect $\mu$ is still given by \req{mudef}. From here it is trivial to check that 
\begin{equation}
E= \Gamma\,  \frac{d}{2}\, \frac{T^{d+2}}{\mu^{\frac{d}{2} +1}} \quad\thus  \quad E  = \frac{d}{2}\, P .
\label{deos}
\end{equation}	
 Note that in converting the entropy and Gibbs potential into field theory quantities,
 because of the compact $v$ direction, we introduce a chemical potential $\mu$ that 
 plays no role in the untwisted backgrounds.
 For the
reasons outlined in \sec{rev} we believe that the appropriate
holographic description of these systems is in terms of the asymptotic
pp-wave spacetimes \req{vacmet} 
even though the geometric evaluation of the Gibbs potential and the
entropy of these non-relativistic CFTs, expressed as a function of the horizon radius
$r_+$, leads to answers that are
identical to those obtained for black holes in untwisted \AdS{d+3}.

The thermodynamics we have discussed is for non-relativistic CFTs in
non-compact space. It would be interesting to study these systems in
finite volume, and to see if they exhibit phase transitions like the
Hawking-Page transition that is well known in the AdS case. 
We suspect the answer is yes given
the similarities of the Euclidean action computation \req{sonshell},
and the fact that we obtain these theories by deforming relativistic
gauge theories.  However, if
we start from $\CN =4$ SYM on a compact manifold, say $\Sp^3 \times
\R$, one needs to pick an appropriate light-cone to carry out the
deformation. For the case of $\Sp^3$ one could presumably use the
non-degenerate Hopf fibre direction to define appropriate light-cone
coordinates and twist the theory by the R-symmetry. 

Apart from the intrinsic interest in developing gravity duals to
non-relativistic condensed matter systems, these Galilean CFTs may
 further the recent developments in the
fluid-gravity correspondence \cite{Bhattacharyya:2008jc}. 
Any interacting field theory in an appropriate long-wavelength limit can
be modeled as a hydrodynamic system; recently
ref.\ \cite{Bhattacharyya:2008jc} developed a precise dictionary
between the dynamics of relativistic conformal fluids and
asymptotically AdS black hole solutions. While this construction
provides an interesting avenue to explore the physics of fluid
dynamics in a holographic setting, our knowledge of relativistic
fluids is less well developed than that of non-relativistic
fluids.  There should be more ways of cross-checking holographic duals
of strongly coupled non-relativistic field theories.

The R-charge twisted $\CN=4$ theory discussed in the paper 
provides in the 
hydrodynamic limit 
an example of a two dimensional fluid,
where hydrodynamic features such as turbulence differ qualitatively
from their higher dimensional counter-parts owing to the inverse
cascade phenomenon \cite{Bernard:2000fx}. In the context of the
fluid-gravity correspondence, ref.\
\cite{VanRaamsdonk:2008fp} suggested that one might see qualitative
differences between gravity in different dimensions, based on
qualitative differences in the turbulent regime. 
Precisely because most work on turbulence is done for
non-relativistic fluids, the non-relativistic
system studied here may provide a particularly good playground for exploring
turbulence. It should however be borne in mind that non-relativistic
conformal fluids are also highly compressible,\footnote{It may however be possible to focus on the low lying shear mode to obtain incompressible Navier-Stokes flow. We thank Dam Son for alerting us to this possibility.} owing to the equation
of state \req{eos}, which is a consequence of scale invariance. The models which violate Galilean conformal invariance would thus be better starting points to realize incompressible fluids.

\subsection*{Acknowledgements}
\label{acks}

It is a pleasure to thank Allan Adams,  Rajesh Gopakumar, Sean Hartnoll, Veronika Hubeny, Pavel Kovtun, Don Marolf, John McGreevy, Shiraz Minwalla, Dam Son, Andrei Starinets and Larry Yaffe for very interesting discussions. We would also like to thank Veronika Hubeny and Larry Yaffe for comments on the manuscript. CH
and MR would like to thank the Galileo Galilei Institute for Theoretical
Physics, Firenze for hospitality during the workshop
``Non-Perturbative Methods in Strongly Coupled Gauge Theories'' and
the INFN for partial support. MR and SFR would like to thank the ICMS,
Edinburgh for hospitality during the workshop ``Gravitational
Thermodynamics and the Quantum Nature of Spacetime".  CPH would like to thank the BIRS for hospitality during the workshop ``Emerging Directions in String Theory''. 
MR would also like to thank the Tata Institute of Fundamental Research, Mumbai, for wonderful hospitality during  the ``Monsoon Workshop on String Theory" and would like to thank the International Center for Theoretical Sciences, India for hospitality and support. MR and SFR are partially supported by STFC.  CPH was supported in part by the NSF under Grant No.\ PHY-0756966. 

\appendix
\section{Anisotropic Galilean field theories}
\label{fluxbgs}

The discussion in the text has been confined to spacetimes of the form
\req{vacmet} with $\nu =1$ supported by matter that preserves the
rotational invariance in the spatial directions. However, it is just
as easy to construct spacetimes where the rotational invariance is
broken by a simple generalization of the construction in
\cite{Son:2008ye}. We would like to obtain the metric (related to
\req{vacmet} by $r =1/z$)
\begin{equation}
ds^2 = \frac{1}{z^2} \, \left(-2 \, du\, dv + d{\bf x}_d + dz^2 \right) - \frac{1}{z^{2\,\nu+2}} \, du^2
\label{FHmet}
\end{equation}	
in $d$-spatial dimensions as  the solution for some equations of motion with some appropriate matter. The matter stress tensor supporting the solution is 
\begin{equation}
T_{uu} \propto \frac{1}{z^{2\,\nu+2}}.
\label{FHuu}
\end{equation}	

In \cite{Son:2008ye} this stress tensor was modeled by a massive
vector. However, we can do just as well with a pair of $p$-form
fields. Let us consider the action
\begin{equation}
\SS = \int\,d^{d+3}x\, \sqrt{-g}\, \left(R - 2\,\Lambda - \frac{1}{2}\, \mid H_{(p+1)}\mid^2 - \frac{1}{2}\, \mid F_{(d+3-p)}\mid^2 \right)- \nu \int\, B_{(p)} \wedge F_{(d+3-p)},
\label{FHlag}
\end{equation}	
with $H_{(p+1)} = dB_{(p)}$ and $F_{(d+3-p)} = dC_{(d+2-p)}$. An
appropriate source is
\begin{eqnarray}
H_{(p+1)} &=& - \frac{\alpha\, (\nu+p)}{z^{\nu+p+1}}\, dz \wedge du \wedge \omega_{(p-1)} ,\nonumber \\
F_{(d+3-p)} &=&  \frac{\alpha\, (\nu+p)}{ z^{\nu +d+3-p}}\, dz \wedge du \wedge (\star\, \omega)_{(d-p+1)} ,
\label{FHdefs}
\end{eqnarray}	
where $\omega_{p}$ is an arbitrary $p$-form on the spatial $\R^{d}$
(parameterized by ${\bf x}_d$) and $(\star\, \omega)_{(d-p)}$ is its
Hodge dual. It is easy to check that this ansatz satisfies the field
equations coming from \req{FHlag} and provides the appropriate stress
tensor \req{FHuu}. In fact the massive vector theory of
\cite{Son:2008ye} corresponds to choosing the $\omega_{(p-1)}$ on the
spatial sections that appears in the $H$-flux to be a scalar and hence
the dual form to be the top-form on $\R^d$, which naturally gives a
mass term (like in massive Type IIA string theory). In more general
cases, the presence of the fluxes breaks rotational invariance ---
while the metric \req{FHmet} has full $SO(d+1)$ rotational symmetry,
the fluxes break $SO(d+1)$ down to $SO(p) \times
SO(d+2-p)$.

This construction is in fact inspired by the Null Melvin Twist along
the world-volume directions of the D-brane to obtain light-like
non-commutative Yang-Mills theories \cite{Alishahiha:2000pu}. Starting
from the non-extremal D3-brane solution \req{nonextD3}, we find a
solution to the IIB equations of motion (for simplicity twisting only
along $x_1$ --- for a more general twist see the solution given in Eq
(C.6) of ref.\ \cite{Hubeny:2005qu})
\begin{equation}
ds_{str}^2 =e^{2\,\varphi}\, \left[r^2\, \left(-f(r)\, dt^2 + dy^2 + dx_1^2 \right) - \beta^2 \,f(r)\,r^6\, (dt+dy)^2\right]+ r^2\, dx_2^2+ \frac{dr^2}{r^2\, f(r)} +  d\Omega_5^2,
\label{nrnonext}
\end{equation}	
supported by NS-NS ($B_{(2)}$) and RR ($C_{(2)}$ and $C_{(4)}$) fluxes in 10 dimensional Type IIB supergravity,
\begin{eqnarray}
B_{(2)} &=&  \beta\, r^4\, e^{2\, \varphi}\, \left( dt + dy\right) \wedge dx_1 ,\nonumber \\
C_{(2)} &=& -\beta\, r^4
\,\left( dt + dy\right) \wedge dx_2, \nonumber \\
F_{(5)} &=&4\, (1+\star) \, {\rm Vol}(\Sp^5),  \nonumber \\
e^{\phi} &=&\frac{1}{\sqrt{1+\beta^2\,r_+^4}} \nonumber \\
f(r) &= & 1- \frac{r_+^4}{r^4} \ . 
\label{othflds}
\end{eqnarray}	

In the geometry \req{nrnonext} we have $\nu =2$ due to the fact that
$g_{uu} \sim r^6$ and the dilaton is constant. Furthermore, the
$\Sp^5$ is also undeformed, unlike the situation encountered in
\req{nonextsol}. Reducing \req{nrnonext} on the $\Sp^5$ we find an
effective 5-dimensional action of the general form \req{FHlag}
\begin{eqnarray}
\SS_{5deff} &=& \frac{1}{16\pi G_5} \int d^5 x \sqrt{-g} \left[
  e^{-2\,\varphi} (R - 2 \Lambda - \frac{1}{12} H_{\mu\nu\lambda}
  H^{\mu\nu\lambda}) - \frac{1}{12} F_{\mu\nu\lambda}
  F^{\mu\nu\lambda} \right] \nonumber \\
&& + \frac{1}{4\pi G_5} \int B_2 \wedge F_3 \ .
\label{5deffact}
\end{eqnarray}	
The five dimensional solution  is then given by the fields
\begin{eqnarray}
ds^2_{str} &=& e^{2\,\varphi} \,\left[ r^2 \,(-f(r)\, dt^2 + dy^2 + dx_1^2) - \beta^2 \,f(r)\, r^6\, (dt + dy)^2 \right] + r^2 \,dx_2^2 + \frac{dr^2}{r^2\, f(r)}, \nonumber \\
B_{(2)} &=& \beta \,r^4 \, e^{2\varphi} (dt + dy) \wedge dx_1 \ , \nonumber \\
C_{(2)} &=& - \beta \, r^4 (dt+dy)\wedge dx_2, \nonumber \\
e^{\varphi} &=& \frac{1}{\sqrt{1+r_+^4\, \beta^2}},
\label{5dmetan}
\end{eqnarray}
for a choice of the cosmological constant
\begin{equation}
\Lambda = -6 - 4 \,r_+^4 \,\beta^2 \,e^{2\, \varphi}
\label{5dconstan}
\end{equation}	
which depends non-trivially on the deformation parameter $\q
=\beta^2\,r_+^4$. This solution is the gravitational dual to
light-like non-commutative $\CN =4$ super Yang-Mills and it would be
interesting to study this example further to understand aspects of
non-relativistic dynamics.

\providecommand{\href}[2]{#2}\begingroup\raggedright\endgroup

\end{document}